\documentclass[preprint,amsmath,amssymb,aps,doublespace]{revtex4}
\usepackage[dvips]{graphicx}
\usepackage{setspace}
\usepackage{amsmath}
\usepackage{amsfonts}
\usepackage{amssymb,color}
\usepackage{subfigure}
\usepackage{graphicx,color}
\usepackage{ifpdf}
\usepackage[latin1]{inputenc}

\begin{document}

\title{Continuously variable spreading exponents in the absorbing Nagel-Schreckenberg model}

\author{
Ronald Dickman,\footnote{email: dickman@fisica.ufmg.br}
}

\address{
Departamento de F\'{\i}sica and
National Institute of Science and Technology for Complex Systems,\\
ICEx, Universidade Federal de Minas Gerais, \\
C. P. 702, 30123-970 Belo Horizonte, Minas Gerais - Brazil
}

\date{\today}

\begin{abstract}
I study the critical behavior of a traffic model with an absorbing state.
The model is a variant of the Nagel-Schreckenberg (NS) model, in which
drivers do not decelerate if their speed is smaller than their headway, the number of empty sites between them and the car ahead.
This makes the free-flow state (i.e., all vehicles traveling at the maximum speed, $v_{max}$, and with all headways greater than $v_{max}$) {\it absorbing}; such states are possible for
for densities $\rho$ smaller than a critical value $\rho_c = 1/(v_{max} + 2)$.
Drivers with nonzero velocity, and with headway equal to velocity, decelerate with probability $p$. This {\it absorbing Nagel-Schreckenberg} (ANS) model, introduced in 
[Phys. Rev. E {\bf 95}, 022106 (2017)], exhibits a line of continuous
absorbing-state phase transitions in the $\rho$-$p$ plane.  Here I study the propagation of activity from a localized seed, and find that the active cluster is compact, as is the active region at long times, starting from uniformly distributed activity.  The critical exponents $\delta$ (governing the decay of the survival probability) and $\eta$ (govening the growth of activity) {\it vary continuously} along the
critical line.  The exponents satisfy a hyperscaling relation associated with compact growth.
\end{abstract}

\maketitle


\section{Introduction}

The Nagel-Schreckenberg (NS) model plays an important role in traffic modeling via cellular automata \cite{ns}.
This simple model represents the effect of fluctuations in driving behavior via the spontaneous reduction of drivers' velocities with probability $p$.
In the original NS model any vehicle with nonzero velocity is subject to this process.
Recently, a modified NS model was introduced in which there is no deceleration
when the velocity is smaller than the headway, that is, the number of empty sites between a vehicle and the one ahead of it \cite{ANS1}. This seemingly minor modification makes the
the free-flow state (all vehicles traveling at the maximum speed, $v_{max}$, and all headways greater than $v_{max}$) {\it absorbing}.  Absorbing free-flow states are possible for densities $\rho$ smaller than a critical value, $\rho_c = 1/(v_{max} + 2)$.

The modified model, called the absorbing Nagel-Schreckenberg (ANS) model, exhibits a line of
absorbing-state phase transitions between free and congested flow in the $\rho-p$ plane \cite{Wangnote}. A key conclusion of Ref.~\cite{ANS1} is that the original NS model possesses a phase transition only for $p=0$.  For $p>0$ there is a source of activity, hence no
absorbing state.  Remarkably, the phase diagram of the ANS model is reentrant, with activity
restricted to intermediate values of $p$ over a certain range of densities.   
The initial study of the ANS model left a number of questions open regarding its
phase diagram and critical behavior.  Here I examine these issues in greater detail, focussing on the propagation of activity starting from a localized active region in an inactive (absorbing) background.  In studies of absorbing-state phase transitions \cite{marro,odor07,henkel}, spreading studies are useful in determining the phase diagram and the critical exponents associated with spread of activity (``spreading exponents").  I find rather surprisingly that the spreading exponents $\delta$ and $\eta$ vary continuously on the boundary between absorbing and active phases.  They satisfy, to good approximation, a hyperscaling relation associated with compact growth \cite{hypscmpt}, consistent with the observation that activity is restricted to a compact region in the ANS model.

The remainder of this paper is organized as follows. In the next section I define the ANS model.  Sec. III reports simulation
results for the phase boundary and spreading exponents, followed by a summary and discussion in Sec. IV.

\section{model}

The NS model and its absorbing counterpart (ANS) are defined on a ring of $L$ sites, each of which may be empty or occupied by a vehicle with
velocity $v=0,1,...,v_{max}$.  (I use $v_{max}=5$, as is standard in studies of the NS model.)
The dynamics, which occurs in discrete time, conserves the number $N$ of vehicles; the associated intensive control parameter is $\rho = N/L$.  Denoting the position of the $i$-th vehicle by $x_i$, we define the headway $d_i =x_{i+1}-x_{i}-1$ as the number
of empty sites between vehicles $i$ and $i+1$.  Each time step consists of four substeps:

 \begin{itemize}
 \item Each vehicle with $v_{i}<v_{max}$ increases its velocity by one unit: $v_{i}\rightarrow v_{i}+1$
 \item Each vehicle with $v_{i}>d_{i}$ reduces its velocity to $v_{i}=d_{i}$.
 \item NS model: each vehicle with $v_i > 0$ reduces its velocity by one unit with probability $p$.\\
 $\;$ ANS model: each vehicle with $v_i > 0$ and velocity equal to headway ($v_i \!=\! d_i$) reduces its velocity by one unit with probability $p$.
 \item All vehicles advance their position in accord with their velocity.
 \end{itemize}

Given the velocities $v_i$ and headways $d_i$, there is no need to keep track of positions:
the final substep is simply $d_i \to d_i - v_i + v_{i+1}$ for $i=1,...,N-1$, and $d_N \to d_N - v_N + v_1$.  Thus $\sum_{i=1}^N d_i$ is also conserved. By the {\it configuration} we understand the set of velocities $\{v_i\}$ and headways $\{d_i\}$.  (Given the configuration, the vehicle positions are defined up to a common additive constant.)  Note that vehicle $i$ has no influence on vehicle $i+1$ ahead of it, but is influenced by this vehicle, since $v_{i+1}$ affects the headway $d_i$, which in turn influences $v_i$.  This implies that changes in velocity propagate upstream, i.e., from vehicle $i$ to vehicle $i-1$.

The modification of the third substep leads to several notable changes in behavior compared to the original NS model \cite{ANS1}. The ANS model exhibits a phase transition for general $p$, whereas the NS model has a phase transition only for $p=0$ \cite{ft6,ft7}.
The low-density, absorbing phase has $v_i = v_{max}$ and $d_{i} \geq v_{max}+1, \; \forall i$; in this phase all drivers advance in a deterministic manner, with the flux given by $j=\rho v_{max}$. The configuration is frozen.  In the active state, by contrast, a nonzero fraction of vehicles have $d_i \leq v_{max}$.  For such vehicles, changes in velocity are possible, and the configuration changes with time.  

Activity in the ANS model is associated with vehicles with $v_i < v_{max}$, or with $d_i = v_{max}$, which can suffer a reduction in velocity.
Since the absorbing state is characterized by $v_i = v_{max}, \forall i$, one
might be inclined to define the activity density simply as $\rho_a = v_{max} - \overline{v}$.  
Nevertheless, not all configurations with $v_i = v_{max}, \forall i$ are absorbing: a vehicle
with $d_i = v_{max}$ may reduce its speed to $v_{max} - 1$.  Since such a
reduction occurs with probability $p$, we define the activity density as:

\begin{equation}
\rho_a =  v_{max} - \overline{v} + p \rho_{a,2} \equiv \rho_{a,1} + p\rho_{a,2},
\label{actdens}
\end{equation}

\noindent where $\overline{v}$ denotes the mean velocity and $\rho_{a,2}$ is the fraction of vehicles with $v_i = d_i = v_{max}$.  Similarly, in the spreading studies reported below, we define the instantaneous activity so:

\begin{equation}
 a(t) = \sum_{i=1}^N \left[(v_{max} - v_i) + p \delta_{v_i, v_{max}} \right].
\end{equation}

\noindent Thus the activity is zero if and only if the configuration is absorbing, that is, if 
$v_i = v_{max}$, {\it and} $d_i > v_{max}, \; \forall i$.  
We refer to vehicles with $v_i < v_{max}$ as ``active type-1" and those with $v_i = v_{max}$ and $d_i = v_{max}$ as ``active type-2."

Systematic investigation of the steady-state flux leads to the conclusion that the $\rho$ - $p$
plane can be divided into three regions \cite{ANS1}.  To begin, we recall that for $\rho > 1/7$ and $p>0$, the mean velocity $\overline{v}$ must be smaller than $v_{max}$, so that the activity is nonzero in this regime.
For $\rho \leq 1/7$, absorbing configurations exist for any value of $p$.  There is nevertheless a region with
$\rho <  1/7$ in which activity is long-lived.  In this region, which we call the {\it active phase}, the steady state depends on whether the initial configuration (IC) has little activity (homogeneous) or
much activity (jammed).  Outside the active phase, all ICs evolve to an absorbing configuration; we call this the
{\it absorbing phase}.

A surprising feature of the boundary between active and absorbing phases is that it is
{\it reentrant}: for a given density in the range $0.114 < \rho < 1/7$,
the absorbing phase is observed for both small and large $p$ values,
and the active phase for {\it intermediate values}.  We denote the upper and lower
branches of the phase boundary by $p_+(\rho)$ and $p_-(\rho)$, respectively; they meet at $\rho_{c,<} \simeq 0.114$.

The phase diagram of the ANS model for a given, fixed $p$ (with $0 < p < 1$),
is similar to that of a conserved stochastic sandpile (CSS) \cite{granada,pruessner}.  In the sandpile, there are no absorbing configurations for
particle density $\rho > z_c - 1$, where $z_c$ denotes the toppling threshold; nevertheless, the absorbing-state phase transition
occurs at a density $\rho_c$ strictly smaller than $z_c -1$.  That is, activity survives for densities
between $\rho_c$ and $z_c -1$, despite the existence of absorbing configurations in this interval of densities.  
Similarly, in the ANS model, although absorbing configurations exist for $\rho < 1/7$,
the phase transition occurs at some density smaller than $1/7$, depending on the deceleration probability $p$.  Another feature shared by the ANS model and the CSS is that activity is coupled to a conserved density (that of vehicles or of particles), which remains frozen in regions devoid of activity.

\section{Spreading simulations}
\label{spreading}

I study the ANS model on rings with $N \simeq 10^4 - 2 \times 10^5$ vehicles, in realizations running to a maximum time of $N$ steps.
The density ($\rho \leq 1/7$) is fixed in the initial condition, using headways $d_i$ set as uniformly as possible.  Recall that
the number of vehicles per site is $\rho = 1/(\overline{d} + 1)$.  Thus for $\rho = 1/n$ with $n$ an integer, we can simply take
$d_i = n-1$, $\forall i$.  For other densities, the initial $d_i$ take two integer values, arranged in a unit cell that is repeated to form a system of $N$ vehicles.  Initially all vehicles have maximum speed, $v_i = v_m = 5$, except for $v_N = 0$. In the ensuing evolution, this initially stationary vehicle provokes localized activity, since vehicles upstream
are obliged to slow down. Since activity propagates at a speed smaller than one vehicle per time step, setting the maximum time equal the number of vehicles guarantees that activity never reaches the first vehicle.

Following the usual practice in analyses of spreading at absorbing-state phase transitions \cite{torre},  I study the survival probability $P(t)$, and the mean value of the activity $a(t)$,
as well as the mean distance between the first and last active vehicles $R(t)$, as functions of time.  At a continuous phase transition to an
absorbing state \cite{marro,henkel,torre}, these quantities are expected to follow asymptotic power laws,

\begin{align}
P(t) & \propto t^{-\delta} \\
a(t) & \propto t^\eta, \\
R(t) & \propto t^{z/2}
\label{eq:abpnr}
\end{align}

\noindent where $\delta$, $\eta$ and $z$ are critical exponents. The power-law dependence of $P$ and $a$ on time provides a
useful criterion for estimating critical parameter values.
In spreading processes free of drift, such as directed percolation (DP), it is common to study $R^2 (t)$, defined as the mean-square
distance of active sites from the original seed, which is expected to follow $R^2 \propto t^z$.
In a process in which activity propagates with nonzero velocity, as in the
present case, the natural analog of $R^2$ is the mean-square radius of gyration of the set of active sites.  
In one dimension it is convenient to define $R$ simply as the distance (in terms of vehicle index) between the right- and leftmost active vehicles.

Before analyzing the statistics of spreading, it is interesting to examine examples of the spreading process in space-time plots, where ``space'' refers to vehicle index not position on the road.
Inspection of such plots confirms that activity propagates upstream, and reveals that the active region is essentially compact.  This means that although the region between the first and last
active vehicles may include inactive ones, there is no branching into two or more active regions that separate over time; the activity density inside the active region tends to a constant nonzero value as time increases.
Thus critical spreading in the ANS model is quite different from that in directed percolation or the contact process \cite{marro,odor07,henkel} (in which active regions can branch, and have an activity density tending to zero as $t \to \infty$), and rather similar to that of {\it compact} directed percolation (CDP) \cite{essam}, in which the active region contains no inactive sites.

\begin{figure}[h!]
  \centering
  \includegraphics[scale=0.8]{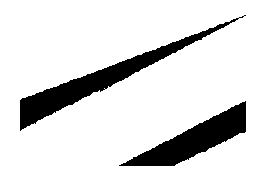}
  \caption{A typical spreading process in the active phase, starting from localized activity.  
  Parameters $\rho=1/7$, $p=0.5$, system size $L=500$, periodic boundaries, maximum time $10^3$.
  Black symbols denote active vehicles.  Horizontal position denotes vehicle number; time increases downward.}
  \label{spr75}
\end{figure}

An example of spreading from a localized seed within the active phase ($\rho = 1/7$, $p=0.5$) is shown in Fig.~\ref{spr75}.  
The active region remains compact,
initially expanding linearly with time, before saturating at a stationary value, once activity has run through the system.  

\begin{figure}[h!]
  \centering
  \includegraphics[scale=0.8]{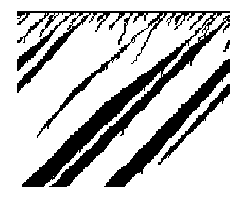}
  \caption{A typical evolution in the active phase starting from uniform activity.  Parameters
  $\rho=1/7$, $p=0.5$, system size $L=200$, maximum time $350$.}
  \label{ev7u}
\end{figure}

Starting from configurations in which activity is distributed uniformly over the entire system, rather than restricted to a localized seed as in spreading experiments, one observes a rapid extinction and
coalescence of activity, until only a single active region remains.   The early stages of this process are illustrated
in Fig.~\ref{ev7u}, for density $\rho=1/7$ and $p=1/2$.  (In this case, initially, even-numbered vehicles have speed zero,
while odd-numbered vehicles have speed $v_{max}$.)
The initially uniform activity breaks into many small active
regions, most of which rapidly die out.  (Although three active regions remain at the latest time shown in the figure,
at later times only a single active region remains.)  The restriction of activity to a single region at long times involves two
processes: disappearance of some active regions via fluctuations, and coalescence of active regions.  The latter be understood qualitatively by noting that the more active a region is, the faster it moves upstream relative to the inactive background,
so that it eventually merges with other less active regions.
A consequence of this spatial segregation of activity is that the ANS model
does not possess a statistically homogeneous active steady state, a feature that sets it apart from many other models exhibiting a phase transition to an absorbing state, such as DP, the contact process (even with particle drift) or
(nondriven) conserved sandpiles.  (It also explains the failure of efforts to devise a
mean-field description based on the hypothesis of a uniform activity density \cite{mfnote}.)

\begin{figure}[h!]
  \centering
  \includegraphics[scale=0.7]{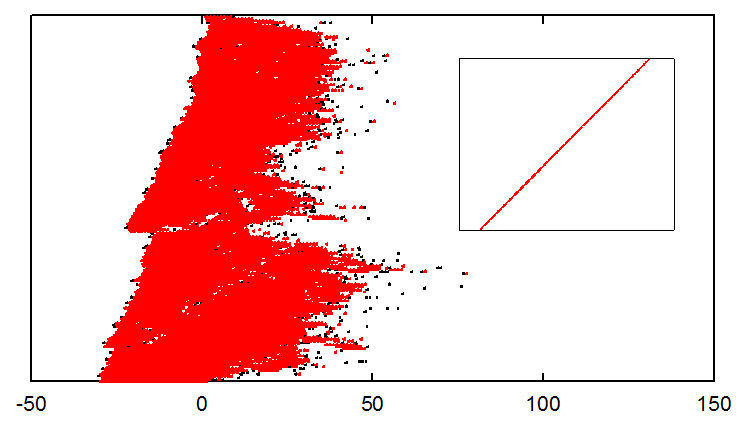}
  \caption{Typical spreading process on the lower critical line, $\rho=1/8$, $p=0.2683$, system size $L=20\,000$,
  maximum time $t_m = 20\,000$.  Red and black symbols
  denote active vehicles of types 1 and 2, respectively.  Inset: horizontal position denotes vehicle number; time increases downward.  Main graph: activity viewed in comoving frame (see text for details).}
  \label{evol3}
\end{figure}

\begin{figure}[h!]
  \centering
  \includegraphics[scale=0.7]{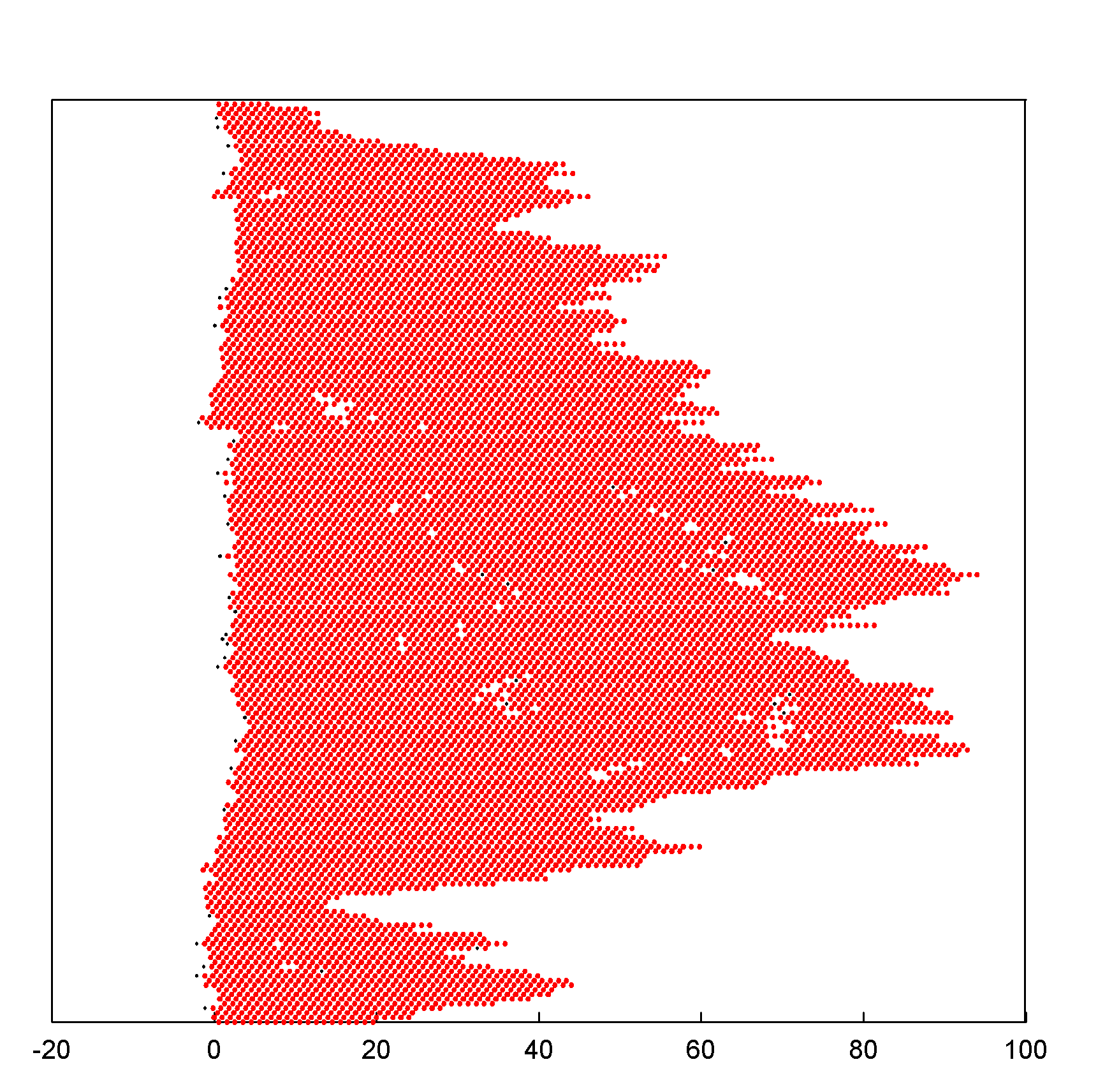}
  \caption{Typical spreading process on the upper critical line viewed in comoving frame
  as in Fig.~\ref{evol4}. Density $\rho=1/8$, $p=0.89595$, system size $L=20\,000$,
  maximum time $t_m = 20\,000$.}
  \label{evol4}
\end{figure}

Turning to examples of spreading along the critical line, I find that propagation of activity in a critical system follows the
same general tendency to compactness noted above, although in this case vehicles may on occasion remain active for some time after the wave of activity has passed.
Figure \ref{evol3} shows spreading on the lower critical line ($\rho=1/8$, $p=0.26830$).  The inset of this figure plots the active
vehicles versus time (increasing downward), as in Figs.~\ref{spr75} and \ref{ev7u}.  At this scale one observes steady upstream propagation of
activity.  To visualize the evolution of the active cluster, the main graph shows the
active vehicles in the comoving frame, i.e., if vehicle $i$ is active a time $t$, a mark is placed at $x = i + ut$, with $-u$ the mean velocity of activity propagation.  (In this case, $u = 0.71196$ vehicles per time step.)
In this figure, active vehicles of types 1 and 2 are denoted in red (black), respectively.  Only a small fraction of
active vehicles are type-2; they are found principally at the boundaries of the active region.  Another notable feature of the evolution is that the trailing (right) edge of the active cluster fluctuates much more than the leading edge.  The latter advances at a fairly steady pace, except for sudden retreats that seem to be associated with the arrival of {\it inactive} clusters (white voids in the field of red) at the edge.
Spreading on the upper critical line (see Fig.~\ref{evol4}) is qualitatively similar.  

\begin{figure}[h!]
  \centering
  \includegraphics[scale=0.7]{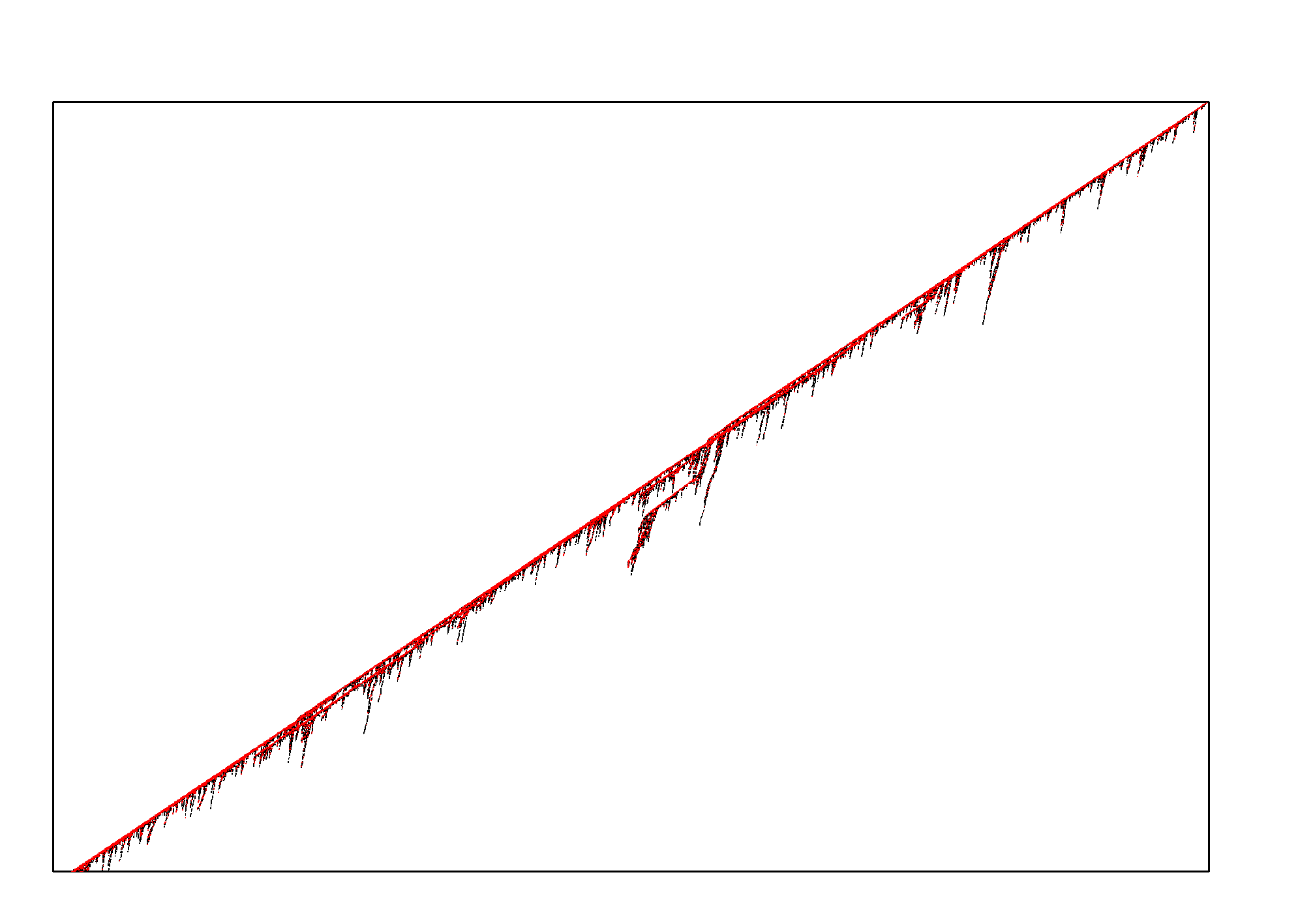}
  \caption{Typical spreading process on the lower critical line, $\rho=0.142518$, $p=0.08148$, $L=t_m=10^5$, as in Fig.~\ref{spr75}. Red and black symbols
  denote active vehicles of types 1 and 2, respectively. }
  \label{evol1}
\end{figure}

The characteristics of spreading change as one approaches density $1/7 \simeq 0.142857$ along the lower critical line.  This regime features smaller values of the deceleration probability $p$ than elsewhere
along the critical line.  Figure \ref{evol1}, for $\rho = 0.142518$ and $p=0.08148$, shows that here, type-2 active vehicles are the majority, and that such vehicles may remain active for a considerable time interval, even when the leading edge of the active region has passed.  Despite this, there is no significant branching of activity on large time scales.  A similar evolution occurs at slightly smaller densities, as shown in Fig.~\ref{evol1}.  In this case the
propagation speed appears to switch between two values.

\begin{figure}[h!]
  \centering
  \includegraphics[scale=0.7]{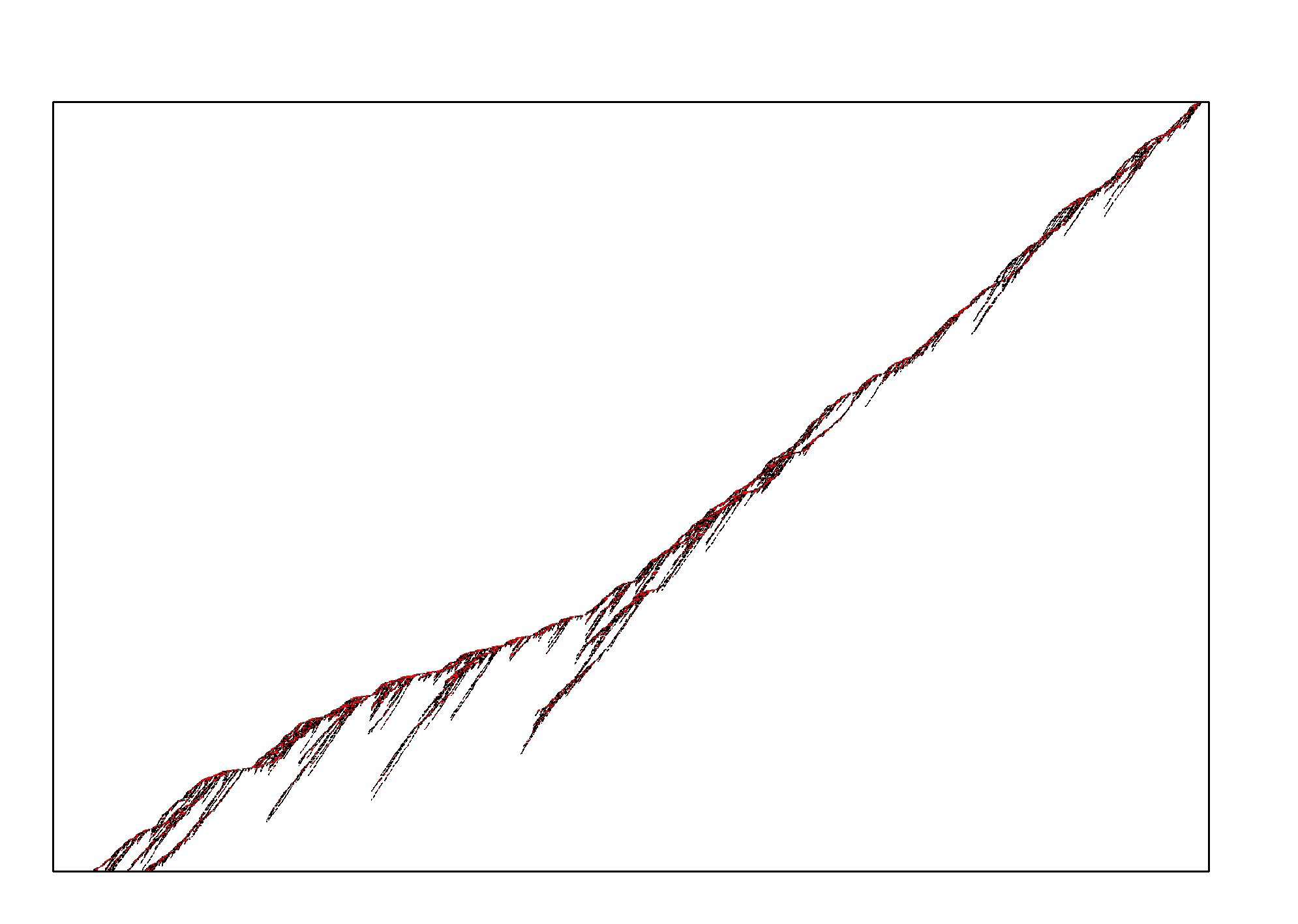}
  \caption{Typical spreading process on the lower critical line, $\rho=0.137931$, $p=0.1301$, $L=t_m=10^4$, as in Fig.~\ref{spr75}. Red and black symbols
  denote active vehicles of types 1 and 2, respectively. }
  \label{evol2}
\end{figure}

\subsection{Phase boundary}

\begin{figure}[h!]
  \centering
  \includegraphics[scale=0.6]{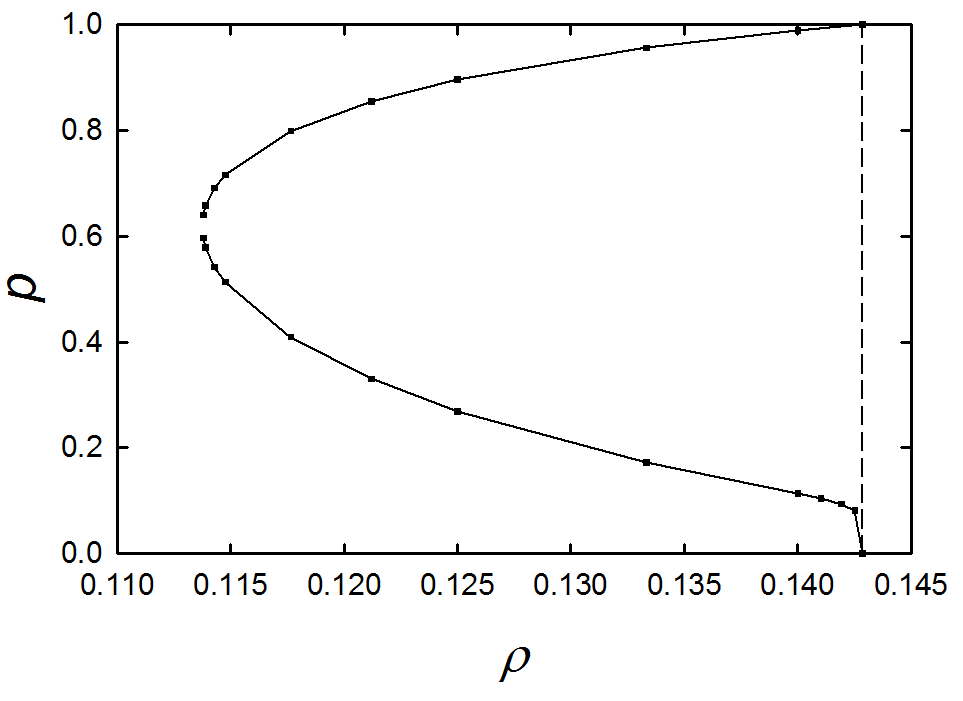}
  \caption{Phase boundary in the $\rho$ - $p$ plane. Error bars smaller than symbols.  Lines are guides to the eye.  The vertical dashed line corresponds to density 1/7.}
  \label{phsbdry}
\end{figure}

Using the criterion of power-law decay of the survival probability $P(t)$, I determine the
phase boundary in the $\rho$ - $p$ plane.  Obtaining reliable estimates for the critical values
$p_<(\rho)$ and $p_> (\rho)$ requires simulations of fairly large systems (typically, a 
maximum size of 1 - 2 $\times 10^5$) and large samples, on the order of $10^7$ ($10^9$)
on the lower (upper) critical line.  The results, plotted in Fig.~\ref{phsbdry}, show the two branches meeting at a density near 0.11375.  The phase boundary appears smooth, but with a
rather sudden downward swerve near the terminus at $\rho=1/7$, $p=0$.

\subsection{Spreading exponents}

\begin{figure}[h!]
  \centering
  \includegraphics[scale=0.6]{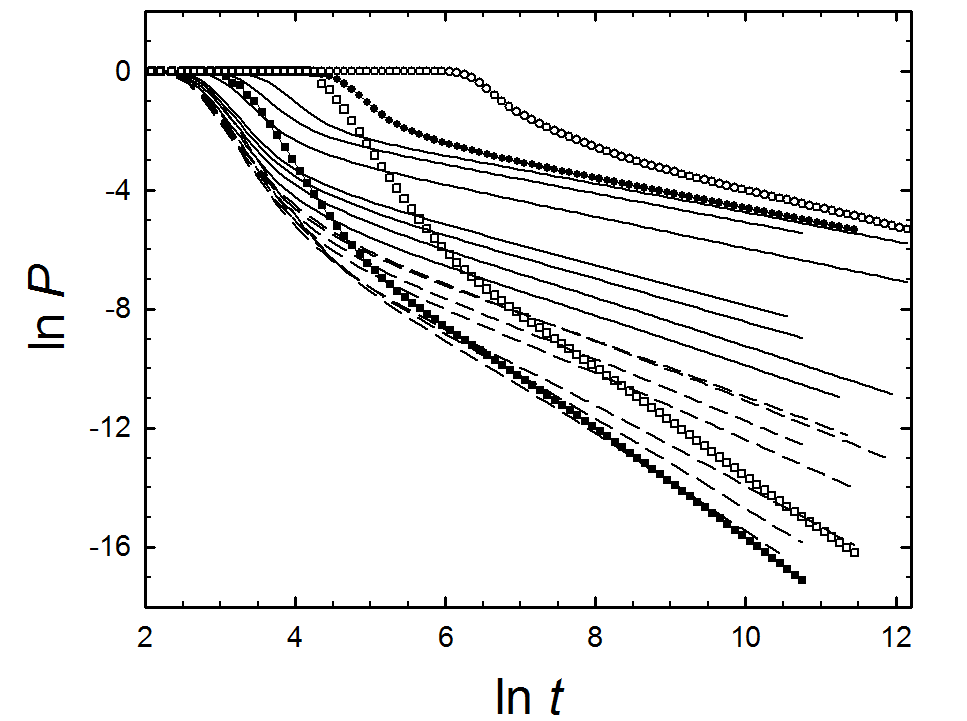}
  \caption{Survival probability $P(t)$ versus time for various points ($\rho, p$) along the phase boundary.    
Solid lines in order of more negative to less negative slope: $\rho$ = 0.113778, 0.113789, 0.11428, 0.11475, 0.117647, $0.\overline{12}$ and 0.125 on the lower critical line.  Broken lines in order of less negative to more negative slope follow the same sequence on the upper critical line.  Filled circles (squares): $\rho = 0.1\overline{3}$, upper (lower) line; open circles
(squares): $\rho=0.14$, upper (lower) line. 
}
  \label{survall}
\end{figure}

The critical exponent $\delta$ governs the asymptotic slope of the survival probability $P(t)$ plotted versus time on log scales; Fig.~\ref{survall} shows this slope varying significantly, and systematically on the phase boundary. The figure reveals three stages in the evolution of $P(t)$: an early stage with
$P(t) = 1$, an intermediate crossover regime, and a final regime of power-law decay.  The first two stages are prolonged near the limiting density of 1/7.

\begin{figure}[h!]
  \centering
  \includegraphics[scale=0.6]{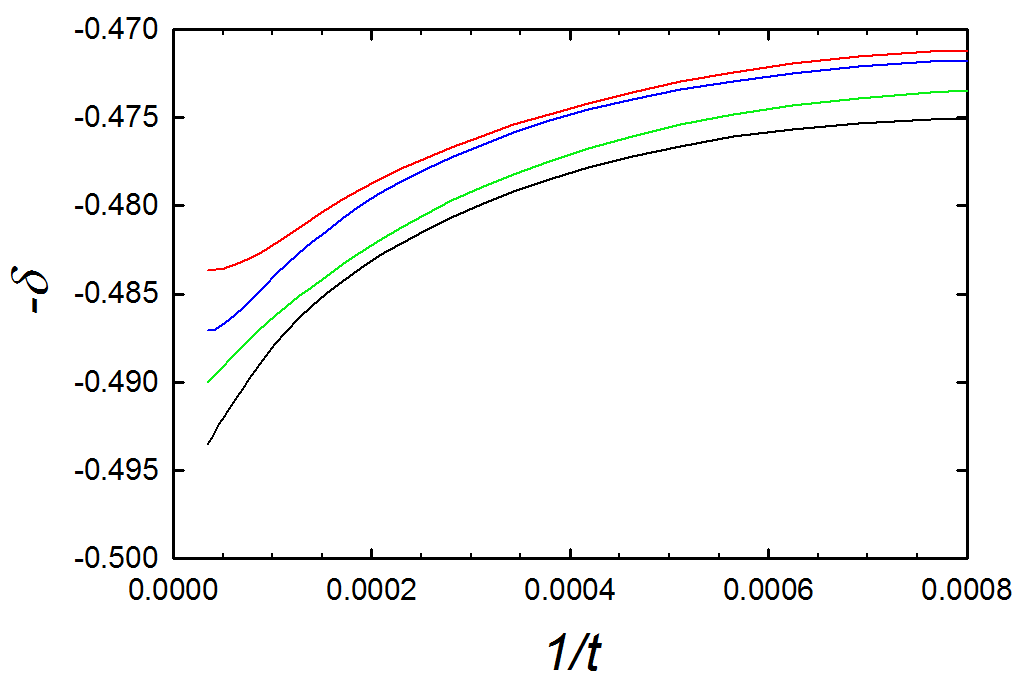}
  \caption{Local-slope plot for $\rho=0.125$  and (lower to upper $p=0.26828$, 0.26829, 0.26830 and 0.26831.
}
  \label{ls125}
\end{figure}

\begin{figure}[h!]
  \centering
  \includegraphics[scale=0.6]{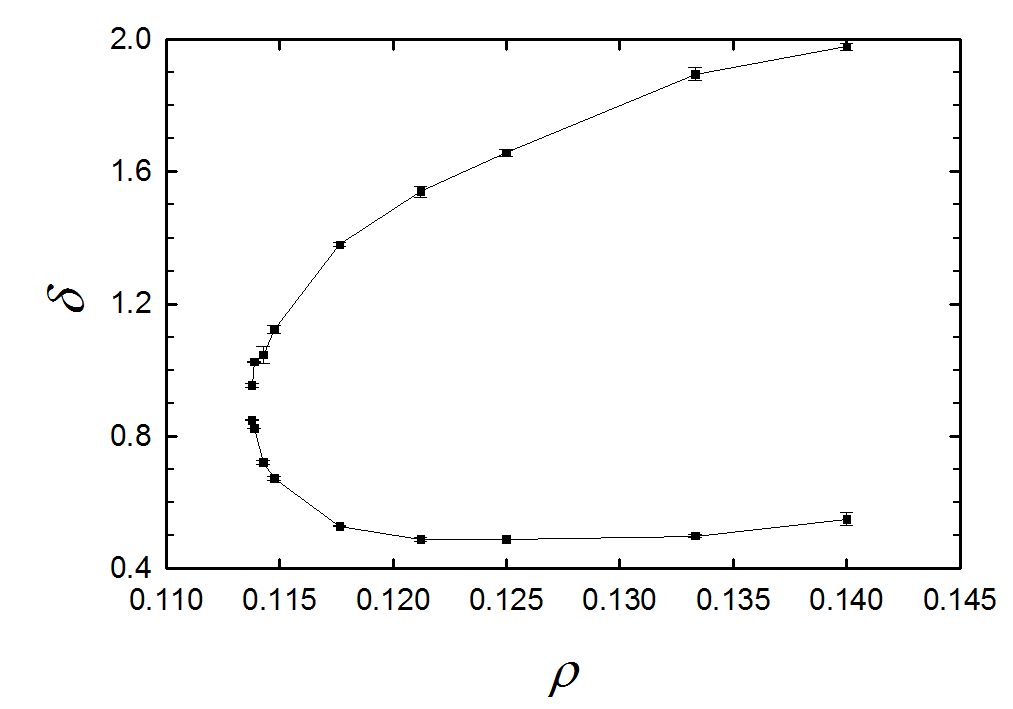}
  \caption{Critical exponent $\delta$ vs. density $\rho$ on the phase boundary.
Lines are guides to the eye.}
  \label{deltaall}
\end{figure}

 To obtain precise results for the spreading exponents and the phase boundary (shown in Fig.~\ref{phsbdry}), I analyze the local slope,
 $\delta(t)$, determined via piecewise linear fits to $\ln P$ versus $\ln t$ on a sliding window
 extending from $\ln (t/3)$ to $\ln 3t$.  Plotting $\delta(t)$ versus $1/t$ allows one to
 eliminate off-critical values and to extrapolate the slope to infinite time, thereby
reducing finite-time corrections to scaling \cite{torre,marro}; an example is shown in Fig.~\ref{ls125}.  Analyses of this kind lead to the results listed in Table I.
 Over most of the phase boundary, $\delta$ increases with $p$; it varies by a factor of four on the critical line.
Figure \ref{deltaall} shows that $\delta$ varies smoothly along the phase boundary,  attaining a maximum value of 1.978 for $\rho=0.14$ on the upper line, and a minimum of about 0.49 at density $0.\overline{12}$ on the lower line.  (Even higher values may occur
on the upper line, for densities between $0.14$ and 1/7.  Exponents have not been determined for this range of densities, due to the long times required to attain the asymptotic scaling regime.)

\begin{table}[h]
\begin{center}
\begin{tabular}{|l|c|c|c|c|} \hline
 $\;\;\;\;\; \rho$ & $p_c$  &  $\delta$ &  $\eta$       &  $z/2$       \\ \hline\hline
\multicolumn{5}{|c|}{lower critical line}                                     \\ \hline
 0.113778 & 0.5970(1)   & 0.8485(10) & -0.324(4)   &  0.555(2)   \\ \hline
 0.113879 & 0.5787(2)   & 0.8240(1)   & -0.304(1)   &  0.528(1)   \\ \hline
 0.114280 & 0.5420(2)   & 0.721(1)     & -0.199(1)  &  0.522(1)    \\ \hline
 0.114750 & 0.5135(1)   & 0.672(3)     & -0.151(4)  &  0.525(1)    \\ \hline
 0.117647 & 0.40962(1) & 0.528(3)     & -0.030(4)  &   0.498(1)    \\ \hline
 0.121212 & 0.33025(5) & 0.488(6)     &  0.005(8)   &  0.492(5)   \\ \hline
 0.125      & 0.26830(1)  & 0.4890(1)   &  0.015(4)  &   0.51(2)     \\ \hline
 0.133333 & 0.171988(2) & 0.503(5)   & -0.003(2)   &  0.498(4)    \\ \hline
 0.140      & 0.11390(5)  & 0.55(2)      &  -0.04(2)   &   0.49(1)   \\ 
\hline \hline
\multicolumn{5}{|c|}{upper critical line}                                     \\ \hline
 0.113778 & 0.6402(3)   & 0.953(7)   & -0.439(8)   &  0.521(1)   \\ \hline
 0.113879 & 0.65766(1) & 1.025(2)   & -0.509(1)   &  0.525(1)   \\ \hline
 0.114280 & 0.6910(5)   & 1.045(25) & -0.524(1)   &  0.529(1)    \\ \hline
 0.114750 & 0.7160(1)   & 1.122(12) & -0.597(10)  &  0.521(3)    \\ \hline
 0.117647 & 0.79867(2) & 1.379(6)   & -0.862(9)    &  0.523(3)    \\ \hline
 0.121212 & 0.85495(3) & 1.540(17) & -1.009(20)  &  0.535(5)   \\ \hline
 0.125      & 0.89595(1)  & 1.657(11) & -1.139(13)  & 0.515(2)     \\ \hline
 0.133333 & 0.95593(3) & 1.895(20) &  -1.392(20) & 0.525(2)    \\ \hline
 0.140      & 0.98853(5)  & 1.978(1)   & -1.463(4)   & 0.510(4)   \\ 
\hline \hline
\end{tabular}
\end{center}
\label{tb:sp}
\caption{Values of spreading exponents along the phase boundary.  }
\end{table}

The critical exponent $\eta$ also varies significantly along the phase boundary, whereas $z/2$ remains in a restricted range centered on 0.515, as shown in Table I.  Rather than plot $\eta$ itself, it seems more informative to consider $\delta + \eta$,
the exponent governing the growth of activity in {\it surviving} realizations.  A hyperscaling
relation for compact growth in $d$ dimensions \cite{hypscmpt}:

\begin{equation}
\delta + \eta = \frac{dz}{2},
\label{hypsc}
\end{equation}

\noindent is pertinent here.  Over much of the phase boundary, $\delta + \eta$ is approximately equal to $z/2$, as expected for compact growth of the active region (see Fig.~\ref{hypsc}).  Over most of the boundary, $\delta+\eta$ and $z$ show little variation: for the eight densities in the interval $[0.113879, 0.14]$ on the lower critical line, I find mean values of 0.509(4) and 0.508(6) for $\delta+\eta$ and $z/2$, respectively; on the upper critical line, the corresponding means are 0.518(3) and 0.523(3).  Departures from $\delta+ \eta = z/2$ may signal that the simulations have not attained the asymptotic regime.  (Note that points with $z/2 < \delta + \eta$ are suspect: asymptotically, the activity cannot grow faster than the size of the active region.)  Longer
simulations and larger sample sizes will be necessary to determine the spreading exponents for $\rho > 0.14$, and to resolve the issue of possible violation
of compact hyperscaling, a task I defer to future work. The present data nevertheless support the hyperscaling relation for compact growth to good precision.

\begin{figure}[h!]
  \centering
  \includegraphics[scale=0.6]{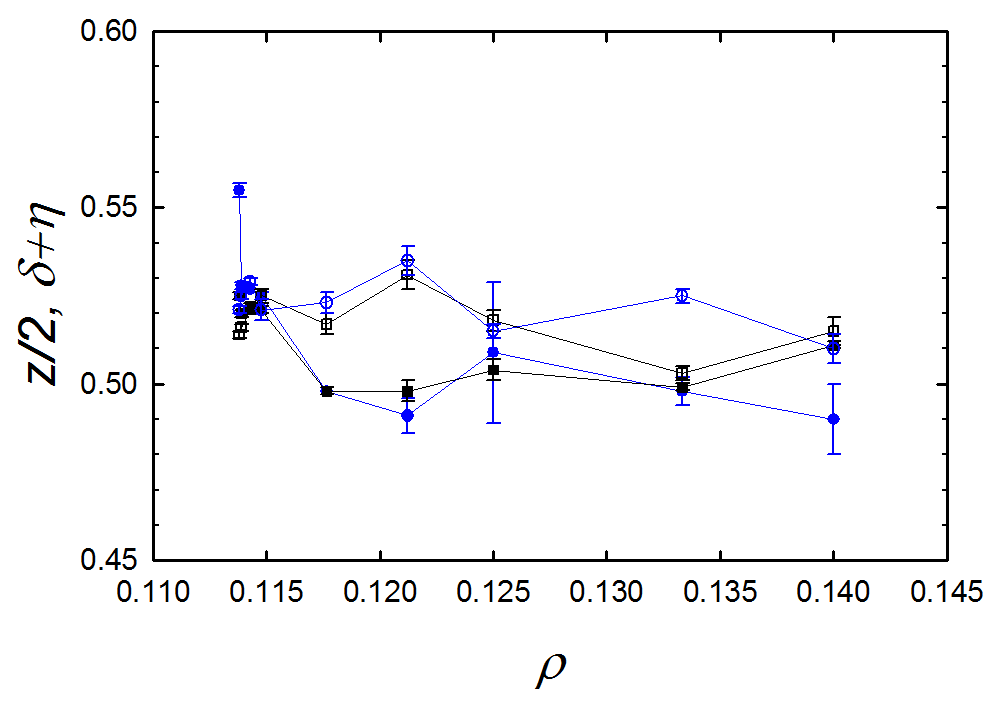}
  \caption{Critical exponents $\delta+\eta$ (squares) and $z/2$ (circles) vs. density $\rho$ on the upper and lower critical lines (open and filled symbols, respectively).
Lines are guides to the eye.}
  \label{hypsc}
\end{figure}

\section{Summary and discussion}

I study the spread of activity from an initial seed in the ANS model, a version of the Nagel-Schreckenberg model that possesses a line of continuous absorbing-state phase transitions in the $\rho$-$p$ plane.  The present study furnishes precise results for the phase boundary, verifies the reentrant nature of the phase diagram and shows that the spreading exponents $\delta$ and $\eta$ vary continuously along the critical line, satisfying the compact-growth hyperscaling relation $\delta + \eta = dz/2$ to good approximation.

A previous study \cite{ANS1} yielded a set of critical exponents consistent with the values $\beta/\nu_\perp =1/2$, $z_d = 1$ and $\nu_\perp = 2$ for the ANS model.  Here $\beta$ is the order-parameter critical exponent, $\nu_\perp$ controls the growth of the correlation length, and $z_d$ is the dynamic exponent, defined via $t_{surv} \sim N^{z_d}$, where $t_{surv}$ is the mean lifetime of activity in a system of $N$ vehicles at the critical point. 
The ratio $\beta/\nu_\perp$ is determined via the scaling behavior $\rho_a \sim N^{-\beta/\nu_\perp}$ at the critical point, where $\rho_a$ denotes the activity density averaged over the entire system.

These results are consistent with the following scenario for the ANS model.  Activity is confined to a compact region with a {\it nonzero activity density}, $\rho_{act}$, even as one approaches the critical line \cite{rhoact}.  In the active phase,
the size $R$ of the active region has a quasistationary (QS) mean $\langle R \rangle_{QS} \sim N\Delta$, where $\Delta$ is the distance from the critical line \cite{QS}.  Then the QS activity density (over all vehicles) is proportional to $\Delta \rho_{act}$, consistent with $\beta = 1$.
At criticality in a finite system, the active region is a fluctuation with typical size 
$\langle R \rangle_{QS} \sim N^{1/2}$ (hence the exponent ratio $\beta/\nu_\perp = 1/2$) and has a mean lifetime $\tau \sim \langle R^2 \rangle_{QS} \sim N$, so that $z_d = 1$.  Since activity propagates at a finite velocity,
the exponent $\nu_{||}$ governing the divergence of the correlation time diverges in the same manner as the correlation length, i.e., $\nu_{||} = \nu_\perp = 2$, consistent with $z_d = \nu_{||}/\nu_\perp = 1$.  

On the critical line in an unbounded system, starting from a localized seed of activity (the situation in spreading simulations), the size $R(t)$ of the active region follows an unbiased random walk, so that the relation
$\langle R(t) \rangle \sim t^{z/2}$ yields $z = 1$.
Note, by contrast, that in the usual scaling relation, $z = 2/z_d$ \cite{torre,marro,henkel},  $z$  is associated with the growth of the mean-square distance $\langle R^2 \rangle$ of activity from the initial seed.  
In the ANS model at criticality, the active region moves upstream at
a nonzero velocity, so that $\langle R^2 \rangle \propto t^2$, and by this definition, $z$ is indeed 2.  Thus the usual definition of $z$ (based on $\langle R^2 \rangle$), is not useful for characterizing the compact or fractal nature of the active region when activity propagates at a nonzero velocity.

A set of spreading exponents consistent with the scenario described above is $z=1$, $\delta = 1/2$ and $\eta=0$, as found, for example, in compact directed percolation \cite{hypscmpt}.  
The scaling relation $\delta = \beta/\nu_{||}$ yields $\delta = 1/2$ when we insert the values
cited above for $\beta$ and $\nu_{||}$, and using this in Eq.~(\ref{hypsc}) we find $\eta = 0$
for $z=d=1$.
But this is not the only possible consistent set of spreading exponents.  Here, surprisingly, I find that $\delta$ and $\eta$ vary continuously, in a situation reminiscent of one-dimensional random walks with an absorbing boundary at the origin and a moveable reflector \cite{rwmpr} or history-dependent step 
length \cite{rwhds}, or CDP subject to moveable reflectors \cite{cdpmr}.   In these cases the survival exponent $\delta$ varies continuously with a parameter.  
With these examples in mind, I note that the active region in the ANS model, being essentially compact, can also be seen as bounded by random walks.  When such a walk advances or retreats, it changes the local configuration of speeds and headways, which can be expected to alter the likelihood of futher changes in position.  While the analogy with random walks is intuitively appealing, it is not clear that the element of irreversible evolution - on the part of the reflector, in the cases studied in \cite{rwmpr,cdpmr} - is present here.  A mapping from the rather complex dynamics of the ANS model to the simple evolution of CDP with moveable partial reflectors is as yet lacking, and is an interesting subject for future study.  Other issues for future work are: the details of the coalescence process; the nature of switching between
slow and rapid propagation (Fig.~6); construction of an inhomogeneous mean-field theory or continuum description.

The present study is motivated by the desire to understand the rather unusual phase diagram uncovered in Ref.~\cite{ANS1} (and by extension, the phase behavior of the original NS model) rather than by possible applications to traffic on highways.  Do the present results have any implications for highway traffic?  I offer the following speculation.  The parameter $p$ (in the ANS model, the probability of deceleration when the headway just equals the velocity) describes driving habits or culture, or some preprogrammed stochastic behavior in the case of driverless vehicles.  From the point of view of avoiding slowdowns (i.e., of {\it minimizing} activity in the ANS model), $p$ values near zero or near unity are optimal. In addition, localized slowdowns die out more rapidly for $p \simeq 1$.  Thus the more cautious behavior (i.e., always decelerating when velocity equals headway) is also, surprisingly, the one minimizing slowdowns.  Whether real traffic follows this simple tendency is an open question.

\begin{acknowledgments}
I thank M.~L.~L. Iannini for helpful discussions.
This work was supported by CNPq, Brazil, under grant No. 303766/2016-6.
\end{acknowledgments}


\begin{thebibliography}{50}
\bibitem{mfnote} For example, a two-site mean-field approximation yields $\rho_c = 1/7$ independent of $p$, and so does not capture the reentrant phase boundary observed in simulations.
\bibitem{rhoact} For example, for $\rho = 0.125$ on the lower critical line, I find $\lim \rho_{act} = 3.45(1)$ as $p$ tends to its criical value from above.
\end{thebibliography}
\end{document}